\documentclass[12pt]{article}
\usepackage{amssymb,amsmath,comment,mathtools}
\usepackage[dvipdfmx]{graphicx,color}
\usepackage[dvipdfmx]{color,hyperref}
\usepackage{cite}
\usepackage{subfigmat}
\usepackage{braket}
\usepackage{hyperref} 	  		
\usepackage{epsfig}
\usepackage{here}
\usepackage{xcolor}
\hypersetup{
    colorlinks=false,
    citebordercolor=green,
    linkbordercolor=red,
    urlbordercolor=cyan,
}
\usepackage{bm}
\graphicspath{{./Figures/}}

\newcommand{\ba}{\begin{align}}
\newcommand{\ea}{\end{align}}

\def\bea{\begin{eqnarray}}
\def\eea{\end{eqnarray}}
 \makeatletter
\def\alt{\mathrel{\mathpalette\gl@align<}}
\def\agt{\mathrel{\mathpalette\gl@align>}}
\def\gl@align#1#2{\lower.6ex\vbox{\baselineskip\z@skip\lineskip\z@
\ialign{$\m@th#1\hfil##\hfil$\crcr#2\crcr\sim\crcr}}} \makeatother
\renewcommand{\thefootnote}{\fnsymbol{footnote}}
\begin{document}
\begin{flushright}
{\small
NITEP 213}
\end{flushright}
\vspace*{1.0cm}

\vspace*{1.0cm}
\begin{center}
\baselineskip 20pt 
{\Large\bf 
Gauge coupling unification and proton decay\\
via 45 Higgs boson in SU(5) GUT}
\vspace{1cm}

{\large 
Naoyuki Haba${}^{a,b}$, Keisuke Nagano${}^{a,c}$, Yasuhiro Shimizu${}^{a,b}$ \\
and Toshifumi Yamada${}^{d}$
} \vspace{.5cm}

{\baselineskip 20pt \it
${}^{a}$Department of Physics, Osaka Metropolitan University, Osaka 558-8585, Japan \\
${}^{b}$Nambu Yoichiro Institute of Theoretical and Experimental Physics (NITEP),
Osaka Metropolitan University, Osaka 558-8585, Japan\\
${}^{c}$Institute of Science and Engineering, Shimane University, Matsue 690-8504, Japan\\
${}^{d}$Department of Physics, Yokohama National University, Yokohama 240-8501, Japan
}

\vspace{.5cm}

\vspace{1.5cm} {\bf Abstract} \end{center}
We study the gauge coupling unification (GCU) and proton decay in a non-supersymmetric SU(5) grand unified theory (GUT) incorporating a 45 representation Higgs field. Our analysis is based on the assumption that Georgi-Jarlskog-type mass matrices for fermions are responsible for explaining the mass ratio of the strange quark and the muon. We examine the conditions of GCU, taking into account the possibility that certain components of the {\bf 45} Higgs field have masses much smaller than the GUT scale. We have found that to satisfy the GCU conditions, at least two components of the {\bf 45} Higgs field should have such small masses.
We search the parameter space to identify regions where the GCU conditions are satisfied, in the scenarios
 where two or three components of the {\bf 45} Higgs boson are hierarchically light. 
If the colored Higgs component of the {\bf 45} Higgs boson has a mass much smaller than the GUT scale, proton decay via colored Higgs boson exchange can occur with an observably large rate. 
We estimate the mass bounds for the colored Higgs component from the proton decay search at Super-Kamiokande and thereby further restrict the parameter space.\\

\thispagestyle{empty}

\newpage
\renewcommand{\thefootnote}{\arabic{footnote}}
\setcounter{footnote}{0}
\baselineskip 18pt
\section{Introduction}

The standard model (SM) is a successful theory that can describe high energy experiments below $\mathcal{O}(100)$ GeV. However, there are problems such as hypercharge quantization and fermion mass hierarchy, leaving physicists to seek more comprehensive theories that go beyond SM. Grand unified theories (GUTs)~\cite{GUT1}-\cite{GUT6} are promising candidates for physics beyond SM, which embed the SM gauge group into a larger group and address hypercharge quantization. The SU(5) GUT~\cite{GUT3} is captivating for its ability to embed the SM gauge group into a simple group. Moreover, the SU(5) GUT proposes the concept of gauge coupling unification (GCU) and proton decay, providing opportunities for experimental verification.

Despite its initial appeal, the SU(5) GUT encounters troubles. Notably, it predicts identical Yukawa coupling constants for the down-type quarks and the charged leptons, resulting in an inability to explain the observed fermion masses. 
Furthermore, precision experiments at LEP have revealed that three gauge coupling constants cannot achieve GCU at high energies. 
In fact, by assuming a mass hierarchy among the physical components of the {\bf 24}-representation GUT-breaking fields, it is possible to achieve GCU,
 but the predicted proton lifetime falls short of the current bound from the proton decay search experiments at Super-Kamiokande~\cite{Super-Kamiokande:2020wjk}.

To address the troubles in the SU(5) GUT, various extensions have been proposed. One notable proposal is to introduce a {\bf 45}-representation Higgs field~\cite{45-1}-\cite{45-5} to explain the observed fermion masses.  
In that model, the Yukawa coupling with the \textbf{45} Higgs field can reproduce the mass ratio of the strange quark and the muon 
 at the GUT scale~\cite{GeorgiJarlskog}.
Still, in non-supersymmetric (SUSY) SU(5) GUTs, three gauge coupling constants fail to achieve GCU. 
To address the GCU problem, SUSY SU(5) GUT has been proposed~\cite{Sakai,Dimpoulos}. However, the non-detection of SUSY particles at LHC experiments invokes doubts about the existence of SUSY.

In this paper, we consider a non-SUSY SU(5) GUT model that incorporates a {\bf 45} Higgs field.
Moreover, we assume that the {\bf 45} Higgs field couples exclusively to the second generation fermions 
 so that the mass ratio of the strange quark and the muon is reproduced through Georgi-Jarlskog-type mass matrices~\cite{GeorgiJarlskog}.
To explain GCU and satisfy the proton lifetime bounds, we assume that the components of the {\bf 45} Higgs field have a hierarchical mass spectrum~\cite{Haba}. 
We solve the renormalization group equations (RGEs) with various hierarchical mass spectra and find out spectra that lead to successful GCU.

If certain components of the {\bf 45} Higgs field are much lighter than the GUT scale, they leave various experimental signatures. In Ref.~\cite{Goto:2023qch}, the authors studied the effect of light components of the {\bf 45} Higgs on quark/lepton flavor-violating decays. In this paper, our focus is on proton decay through the colored-Higgs component of the {\bf 45} Higgs field.
We study various proton decay channels under the assumption that the {\bf 45} Higgs field couples only to the second generation fermions to explain the strange quark and muon mass ratio. 
We search for parameter regions where GCU is realized and at the same time the current experimental bounds on proton decay are satisfied.

This paper is organized as follows: In Sec.~\ref{Model}, we give a review of a non-SUSY SU(5) GUT with a {\bf 45} Higgs field. In Sec.~\ref{GCU}, we investigate the GCU conditions assuming that the components of the {\bf 45} Higgs field have a hierarchical mass spectrum. In Sec.~\ref{proton decay}, we examine the contribution of the colored-Higgs component of the {\bf 45} Higgs field to proton decay and calculate proton lifetimes. We then identify the parameter region allowed by all the experiments. Sec.~\ref{Conclusion} concludes the paper.


\section{Model: $SU(5)$ GUT with a {\bf 45} Higgs Field\label{Model}}
We review a non-SUSY $SU(5)$ GUT that incorporates a {\bf 45}-representation Higgs field. 
The SM matter fields are embedded into $\bm{\bar{5}}$- and $\bm{10}$-representation fermions in the same way as the original SU(5) GUT, as follows:
\begin{align}
	\bm{\bar{5}}: \overline{\psi}_i^a=
\begin{pmatrix}
	d^{~c}_1 \\
	d^{~c}_2 \\
	d^{~c}_3 \\
	\ell^- \\
	-\nu \\
\end{pmatrix}_{iL},~~~~
\bm{10}:\chi_{iab} 
=\frac{1}{\sqrt{2}}
\begin{pmatrix}
	0 & u^{~c}_3 & -u^{~c}_2 & -u_1 & -d_1 \\
	-u^{~c}_3 & 0 & u^{~c}_1 & -u_2 & -d_2 \\
	u^{~c}_2 & -u^{~c}_1 & 0 & -u_3 & -d_3 \\
	u_1 & u_2 & u_3 & 0 & -\ell^+ \\
	d_1 & d_2 & d_3 & \ell^+ & 0  \\
\end{pmatrix}_{iL}.
\end{align}
where $i=1,2,3$ is the generation index.
Also, the SM gauge fields are embedded into a $\bm{24}$ representation as follows:
\begin{align}
	\bm{24}:A_\mu
	=
	\begin{pmatrix}
	G^1_{~1}-\frac{2B}{\sqrt{30}} & G^2_{~1} & G^3_{~1} & \bar{X}_1 & \bar{Y}_1 \\
	G^1_{~2} & G^2_{~2}-\frac{2B}{\sqrt{30}} & G^3_{~2} & \bar{X}_2 & \bar{Y}_2 \\
	G^1_{~3} & G^2_{~3} & G^3_{~3}-\frac{2B}{\sqrt{30}} & \bar{X}_3 & \bar{Y}_3 \\
	X^1 & X^2 & X^3 & \frac{W^0}{\sqrt{2}}+\frac{3B}{\sqrt{30}} & W^+ \\
	Y^1 & Y^2 & Y^3 & W^- & -\frac{W^0}{\sqrt{2}}+\frac{3B}{\sqrt{30}}  \\
\end{pmatrix},
\end{align}
where we use the same convention as Ref.~\cite{Langacker:1980js}.

Next, we show the Yukawa sector. 
To obtain Georgi-Jarlskog-type fermion mass matrices, we assume that the {\bf 45} Higgs field couples only to the second generation fermions. 
As a result, the {\bf 45} Higgs field does not couple to two {\bf 10}-representation fermions because this coupling is asymmetric in the flavor space.
Thus the Yukawa interactions are given by
\begin{align}
	\mathcal{L}_{Y}=Y_{\bar{5}} \overline{\psi}^a\chi_{ab} \overline{5}_H^b+\frac{1}{8}Y_{10}\varepsilon_{abcde}\chi^{ab}\chi^{cd}5_H^e+Y_{\overline{45}} \overline{\psi}^a\chi_{ab} \overline{45}_H^b+\text{h.c.}
\end{align}

The {\bf 45} Higgs field is decomposed into components in the SM representations as follows:
\begin{align}
    45_H=(1,2)_{\frac{1}{2}}
    \oplus(3,1)_{-\frac{1}{3}}
    \oplus(3,3)_{-\frac{1}{3}}
    \oplus(\overline{3},1)_{\frac{4}{3}}    
    \oplus(\overline{3},2)_{-\frac{7}{6}}
    \oplus(\overline{6},1)_{-\frac{1}{3}}
    \oplus(8,2)_{\frac{1}{2}}.
\end{align}
We assume that these components have a hierarchical mass spectrum to achieve GCU. 
In the next section, we examine the contributions of the GUT particles to the beta functions and how they realize GCU.

\section{Gauge Coupling Unification\label{GCU}}

We study the running of the gauge couplings. By solving the 1-loop RGEs of the gauge couplings, the running gauge couplings at energy scale $Q$ are expressed as follows:
\begin{align}\label{1-loop beta function}
	\alpha_i^{-1}(M_Z)=~\alpha_i^{-1}(Q)-\frac{1}{2\pi}\beta_i \log\biggl(\frac{M_Z}{Q} \biggr),
\end{align}
where $M_Z$ is the $Z$ boson mass and $\beta_i$'s represent the beta functions of the SM gauge couplings.  
We calculate the contributions of the {\bf 45} Higgs field components and other particles to the beta functions using the following formula:
\begin{align}
	\beta_{i}=-\frac{11}{3}C_2(G_i)+\frac{4}{3}\kappa S_2(F_i)+\frac{1}{3}\eta S_2(S_i),
\end{align}
where we use the same convention as Ref.~\cite{Intermediate scale}. 
The beta functions thus calculated are listed in Table~\ref{betacoefiecients}.
\begin{table}[H]
\centering
\caption{Beta functions of the particles in a $SU(5)$ GUT with a {\bf 45} Higgs boson}
\scalebox{1.0}{
  \begin{tabular}{|l|c|c|c|c|} \hline
    & SM Rep & $\beta_3$ & $\beta_2$ & $\beta_1=3/5\cdot\beta_Y$ \\ \hline
    $Q_i=\begin{pmatrix}
      u_i \\
      d_i
    \end{pmatrix}_L$ & $(3,2)_{1/6}$ & $2$ & $3$ & $1/5$ \\
    $u_{i}^{~c}=(u_{i}^{~c})_L$ & $(\overline{3},1)_{-2/3}$ & $1$ & $0$ & $8/5$ \\
    $d_i^{~c}=(d_{i}^{~c})_L$ & $(\overline{3},1)_{1/3}$ & $1$ & $0$ & $2/5$ \\\hline
    $L_i=\begin{pmatrix}
      \nu_{e i} \\
      e_i
    \end{pmatrix}_L$ & $(1,2)_{-1/2}$ & $0$ & $1$ & $3/5$ \\
    $e_i^{~c}=(e_{i}^{~c})_L$ & $(1,1)_{1}$ & $0$ & $0$ & $6/5$ \\ \hline
    $H=\begin{pmatrix}
      \phi_+ \\
      \phi_0
    \end{pmatrix}$ & $(1,2)_{-1/2}$ & $0$ & $1/6$ & $1/10$ \\
    $H^c=\begin{pmatrix}
      H_1 \\
      H_2 \\
      H_3
    \end{pmatrix}$ & $(\overline{3},1)_{1/3}$ & $1/6$ & $0$ & $1/15$ \\ \hline
	$(X^\mu,Y^\mu)$ & $(3,2)_{-5/6}$ & $-7$ & $-21/2$ & $-35/2$ \\ \hline
    $\Sigma_8$ & $(8,1)_{0}$ & $1/2$ & $0$ & $0$  \\
    $\Sigma_3$ & $(1,3)_{0}$ & $0$ & $1/3$ & $0$  \\
    $\Sigma_{24}$ & $(1,1)_{0}$ & $0$ & $0$ & $0$  \\ \hline
    $(1,2)_{-1/2}$ & $(1,2)_{-1/2}$ & $0$ & $1/6$ & $1/10$  \\
    $(\overline{3},1)_{1/3}$ & $(\overline{3},1)_{1/3}$ & $1/6$ & $0$ & $1/15$  \\
    $(\overline{3},3)_{1/3}$ & $(\overline{3},3)_{1/3}$ & $1/2$ & $2$ & $1/5$  \\
    $(3,1)_{-4/3}$ & $(3,1)_{-4/3}$ & $1/6$ & $0$ & $16/15$  \\
    $(3,2)_{7/6}$ & $(3,2)_{7/6}$ & $1/3$ & $1/2$ & $49/30$  \\
    $(6,1)_{1/3}$ & $(6,1)_{1/3}$ & $5/6$ & $0$ & $2/15$  \\
    $(8,2)_{-1/2}$ & $(8,2)_{-1/2}$ & $2$ & $4/3$ & $4/5$  \\ \hline
  \end{tabular}}
  \label{betacoefiecients}
\end{table}
Using Table~\ref{betacoefiecients}, we examine GCU with various mass spectra of the components of the {\bf 45} Higgs field. 
Plugging in these $\beta$ functions, we solve the 1-loop RGEs and obtain the following solutions:
\begin{align}
	\alpha_1^{-1}(M_Z)=\alpha_5^{-1}(\Lambda)+\frac{1}{2\pi}\biggl[&\frac{41}{10}\log\frac{M_Z}{\Lambda}+\frac{1}{15}\log\frac{M_{H_C}}{\Lambda}-\frac{35}{2}\log\frac{M_{XY}}{\Lambda}+\frac{1}{10}\log\frac{M(1,2)}{\Lambda}\nonumber\\
	&+\frac{1}{15}\log\frac{M(\overline{3},1)}{\Lambda}+\frac{1}{5}\log\frac{M(\overline{3},3)}{\Lambda}
	 +\frac{16}{15}\log\frac{M(3,1)}{\Lambda}\nonumber\\
	&+\frac{49}{30}\log\frac{M(3,2)}{\Lambda}
	 +\frac{2}{15}\log\frac{M(6,1)}{\Lambda}
	 +\frac{4}{5}\log\frac{M(8,2)}{\Lambda}\biggr]-\frac{5}{12\pi},\\
	\alpha_2^{-1}(M_Z)-\frac{2}{12\pi}=\alpha_5^{-1}(\Lambda)+\frac{1}{2\pi}\biggl[&-\frac{19}{6}\log\frac{M_Z}{\Lambda}-\frac{21}{2}\log\frac{M_{XY}}{\Lambda}+\frac{1}{3}\log\frac{M_\Sigma}{\Lambda}+\frac{1}{6}\log\frac{M(1,2)}{\Lambda}\nonumber\\
	&+2\log\frac{M(\overline{3},3)}{\Lambda}
	 +\frac{1}{2}\log\frac{M(3,2)}{\Lambda}
	 +\frac{4}{3}\log\frac{M(8,2)}{\Lambda}\biggr]-\frac{5}{12\pi},\\
	\alpha_3^{-1}(M_Z)-\frac{3}{12\pi}=\alpha_5^{-1}(\Lambda)+\frac{1}{2\pi}\biggl[&-7\log\frac{M_Z}{\Lambda}+\frac{1}{6}\log\frac{M_{H_C}}{\Lambda}-7\log\frac{M_{XY}}{\Lambda}+\frac{1}{2}\log\frac{M_\Sigma}{\Lambda}\nonumber\\
	&+\frac{1}{6}\log\frac{M(\overline{3},1)}{\Lambda}
	 +\frac{1}{2}\log\frac{M(\overline{3},3)}{\Lambda}
	 +\frac{1}{6}\log\frac{M(3,1)}{\Lambda}\nonumber\\
	&+\frac{1}{3}\log\frac{M(3,2)}{\Lambda}
	 +\frac{5}{6}\log\frac{M(6,1)}{\Lambda}
	 +2\log\frac{M(8,2)}{\Lambda}\biggr]-\frac{5}{12\pi},
\end{align}
where $M(a,b)$ denotes the mass of the {\bf 45} Higgs field component in $(a,b)$-representation of SM SU(3)$\times$SU(2). From these equations, we derive the following two independent equations:
\begin{align}
	\alpha_1^{-1}(M_Z)-3\alpha_2^{-1}(M_Z)+2\alpha_3^{-1}(M_Z)=-\frac{1}{2\pi}\biggl[&\frac{2}{5}\log\frac{M_{H_C}}{M_Z}-\frac{2}{5}\log\frac{M(1,2)}{\Lambda}+\frac{2}{5}\log\frac{M(\overline{3},1)}{\Lambda}\nonumber\\
	&-\frac{24}{5}\log\frac{M(\overline{3},3)}{\Lambda}
	 +\frac{7}{5}\log\frac{M(3,1)}{\Lambda}
	 +\frac{4}{5}\log\frac{M(3,2)}{\Lambda}\nonumber\\
	&+\frac{9}{5}\log\frac{M(6,1)}{\Lambda}
	 +\frac{4}{5}\log\frac{M(8,2)}{\Lambda}\biggr],
	 \\
	-5\alpha_1^{-1}(M_Z)+3\alpha_2^{-1}(M_Z)+2\alpha_3^{-1}(M_Z)=-\frac{1}{2\pi}\biggl[&42\log\frac{M_{XY}}{M_Z}
	 +2\log\frac{M_\Sigma}{M_Z}
	 +6\log\frac{M(\overline{3},3)}{\Lambda} 
	 -5\log\frac{M(3,1)}{\Lambda} \nonumber\\
	&
	 -6\log\frac{M(3,2)}{\Lambda}
	 +\log\frac{M(6,1)}{\Lambda}
	 +4\log\frac{M(8,2)}{\Lambda}\biggr].
 \label{GCUcondition}
\end{align}
Notice that $\Lambda$ cancels out in the above equations, rendering each equation independent of $\Lambda$.
From these equations, we search for mass spectra of the {\bf 45} Higgs field components that achieve GCU.
To avoid rapid proton decay via $XY$ gauge boson exchange, their mass $M_{XY}$ must be larger than $6.0 \times 10^{15}$ GeV. In addition,
the $(\overline{3},1)$ boson mass must be larger than $\mathcal{O}(10^{10})$ GeV when we consider constraints from proton decay via the $(\overline{3},1)$ boson exchange.

The achievement of GCU and the satisfaction of the proton lifetime bounds is impossible if all the components of the {\bf 45} Higgs field have a degenerate mass.
Hence, we consider scenarios where some components of the {\bf 45} Higgs field have masses hierarchically smaller the other components.
For simplicity, we assume that the latter components as well as the $XY$ gauge bosons,
 the physical components of the {\bf 24} scalar field and the colored components of the {\bf 5} Higgs field
 have a degenerate mass, denoted by $\overline{M}$.

First, we examine the case where only one component is light.
The solution to the GCU conditions Eq.~(\ref{GCUcondition}) exists,
since there are two unknowns, $\overline{M}$ and the light component mass, in two equations. 
However, we find that $\overline{M}$ is larger than the Planck scale and so we discard this solution.

Next, we consider the case where two components are light.
We find that solutions exist only when components $(\overline{3},3)$ and $(8,2)$ are light. This is due to the large beta functions of $(\overline{3},3)$ and $(8,2)$, resulting in a significant modification to the running of the gauge couplings. For other combinations of light components, we observe that $\overline{M}$ is larger than the Planck scale, and so we discard such solutions. 
In Table~\ref{mass spectra1}, we list two examples of allowed mass spectra. The mass spectrum (1) is obtained by assuming that $\overline{M}$ is equal to the lower limit of the $XY$ boson mass, $6.0\times10^{15}$~GeV.
The mass spectrum (2) is obtained by taking the mass of $(8,2)$ to be larger than 1~TeV.
This choice is because we estimate the experimental lower bound of the mass of $(8,2)$ to be 1~TeV,
based on Fig.~7 of Ref.~\cite{Hayreter:2017wra}
that constrains the mass of a neutral color octet scalar using the channel at the LHC where color octet scalars are pair-produced
 and give a pair of di-bottom resonances through their decays.
Although the neutral octet scalar in our model mainly decays into two strange quarks,
 we think that the bound in Fig.~7 of Ref.~\cite{Hayreter:2017wra} is largely applicable to our case.
Notice that the mass of $(8,2)$ would be lower if $\overline{M}$ were increased. Thus, we conclude that $\overline{M}$ should be below $9.0\times10^{15}$ GeV to achieve GCU.

\begin{figure}[H]
    \centering
    \includegraphics[width=80mm]{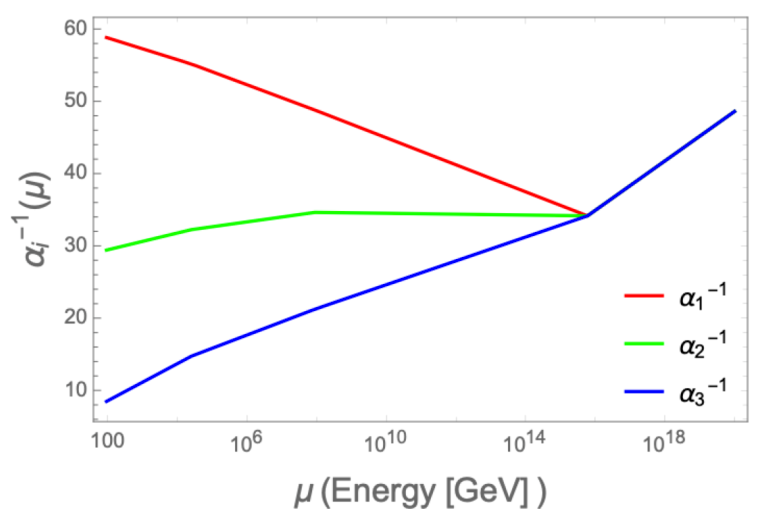} \hspace{0.5cm}
    \includegraphics[width=80mm]{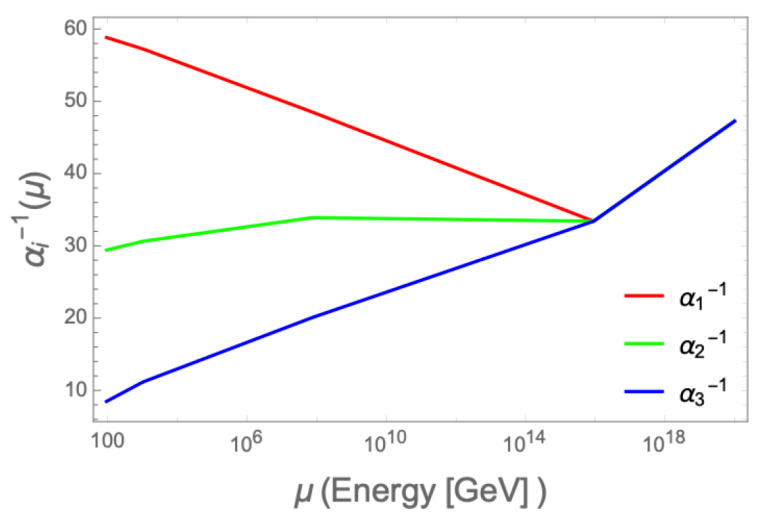}
    \caption{Runnings of the gauge coupling constants that achieve GCU with two light components of the {\bf 45} Higgs field. The left panel is for the mass spectrum (1) and the right panel is for (2).}
    \label{fig:GCU}
\end{figure}

\begin{table}[H]
\caption{Mass spectra of the {\bf 45} Higgs boson components and other GUT particles satisfying the GCU conditions when the $(\overline{3},3)$ and $(8,2)$ components are light.}
\centering
\scalebox{0.75}{
  \begin{tabular}{|l|c|c|c|c|c|c|c|c|c|c|} \hline
  	& $M_{XY}$ & $M_{H_C}$ & $M_{\Sigma}$ & $M(1,2)$ & $M(\overline{3},1)$ & $M(\overline{3},3)$ & $M(3,1)$ & $M(3,2)$ & $M(6,1)$ & $M(8,2)$  \\ \hline
  	$(1)$  & $6.0\times10^{15}$ & $6.0\times10^{15}$ & $6.0\times10^{15}$ & $6.0\times10^{15}$ & $6.0\times10^{15}$ & $8.9\times 10^{7}$ & $6.0\times10^{15}$ & $6.0\times10^{15}$ & $6.0\times10^{15}$ & $2.5\times10^{4}$ \\ \hline
  	$(2)$ & $9.0\times10^{15}$ & $9.0\times10^{15}$ & $9.0\times10^{15}$ & $9.0\times10^{15}$ & $9.0\times10^{15}$ & $7.6\times 10^{7}$ & $9.0\times10^{15}$ & $9.0\times10^{15}$ & $9.0\times10^{15}$ & $1.0\times10^{3}$ \\ \hline 
  \end{tabular}}
  \label{mass spectra1}
\end{table}
\noindent
In Fig.~\ref{fig:GCU}, we illustrate the evolution of the gauge couplings for the mass spectra provided in Table~\ref{mass spectra1}. It is evident from the figures that GCU is attainable due to the presence of the two light components.

\begin{figure}[H]
    \includegraphics[width=55mm]{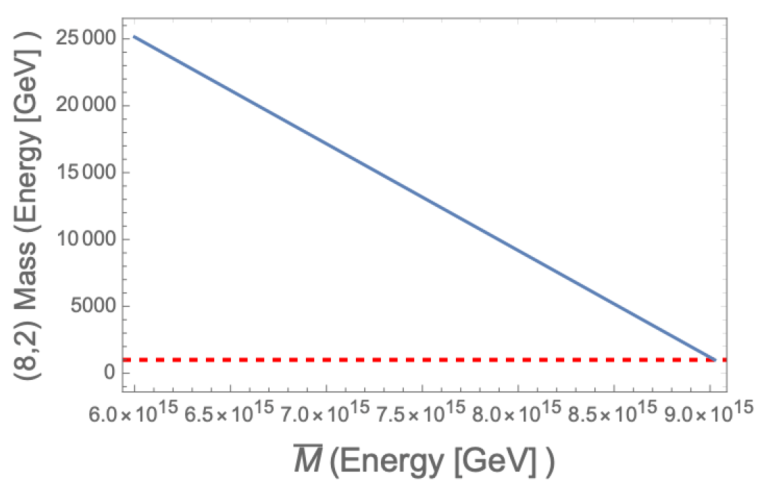}
    \includegraphics[width=55mm]{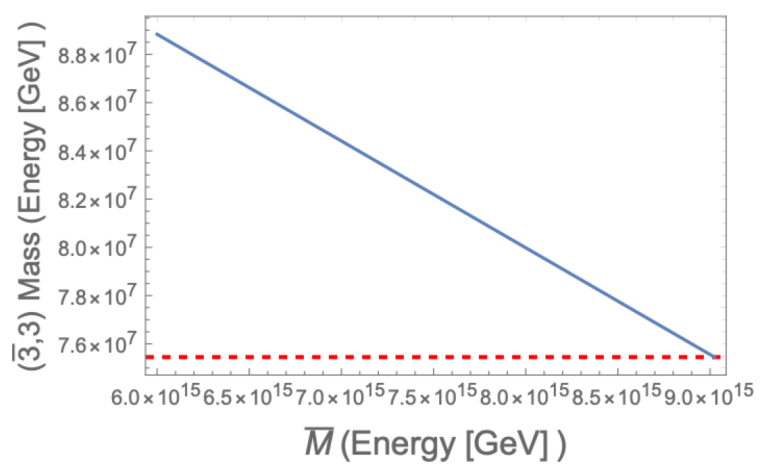}
    \includegraphics[width=55mm]{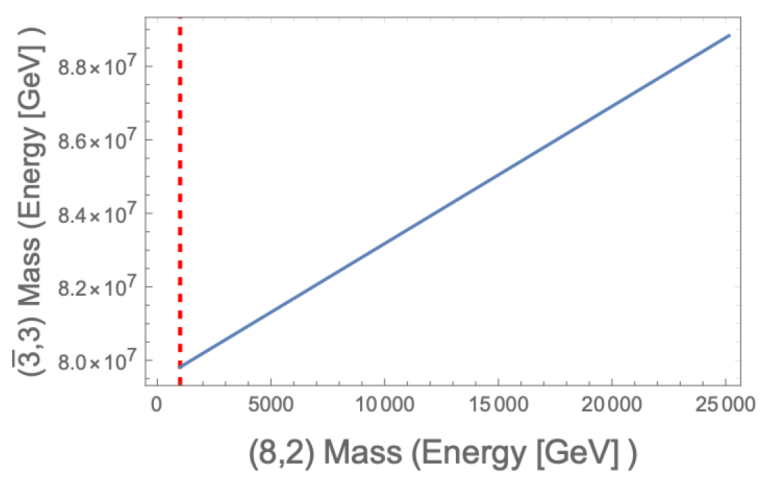}
  \caption{The blue solid lines illustrate the solutions to the GCU conditions. In the left panel, the solutions are plotted on the $\overline{M}-M(8,2)$ plane,
  in the central panel, they are plotted on the $\overline{M}-M(\overline{3},3)$ plane, and in the right panel, they are plotted on the $M(8,2)-M(\overline{3},3)$ plane. The red dashed lines correspond to the constraint from the LHC that the mass of $(8,2)$ be larger than 1~TeV.
  }
  \label{curves}
\end{figure}

In general, there are infinite number of solutions to the GCU conditions Eq.~(\ref{GCUcondition}) in the case where two components, $(\overline{3},3)$ and $(8,2)$, are light, since there are three unknowns, $\overline{M}$, $M(\overline{3},3)$, $M(8,2)$, in two equations. In this case, the solutions satisfying the GCU conditions form a curve on the plane defined by two out of the three unknowns.
In Fig.~\ref{curves}, the solutions of the GCU conditions are plotted on the $\overline{M}-M(8,2)$, $\overline{M}-M(\overline{3},3)$, $M(8,2)-M(\overline{3},3)$ planes, respectively.

Next, we consider the case where three components of the {\bf 45} Higgs field are light. As in the two-light-components case, we find that 
$(\overline{3},3)$ and $(8,2)$ tend to be light to satisfy the GCU conditions. We focus on the case where $(\overline{3},1)$, $(\overline{3},3)$, $(8,2)$ are light, since light $(\overline{3},1)$
 can induce rapid proton decay and so we can obtain constraints not only from the GCU conditions but also from the proton lifetime bounds. We will discuss proton decay through the $(\overline{3},1)$ component in the next section. 
In Table~\ref{mass spectra2}, we list two examples of mass spectra where $(\overline{3},1)$, $(\overline{3},3)$, $(8,2)$ are light to satisfy the GCU conditions.

\begin{figure}[H]
    \centering
    \includegraphics[width=80mm]{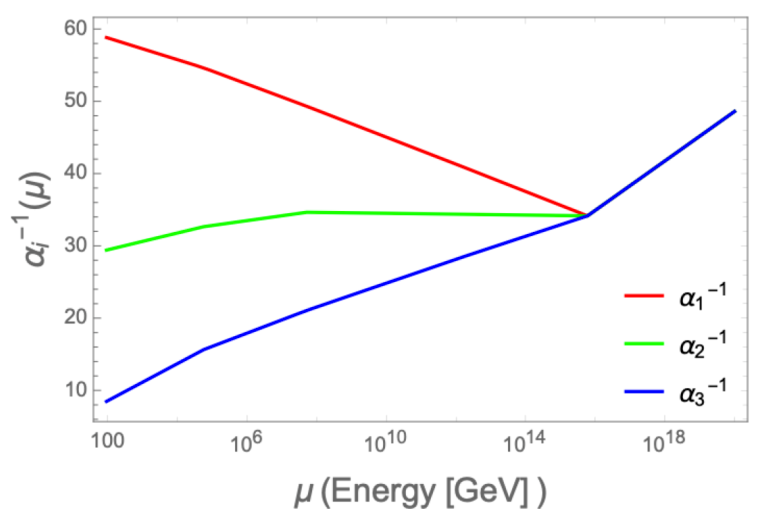} \hspace{0.5cm}
    \includegraphics[width=80mm]{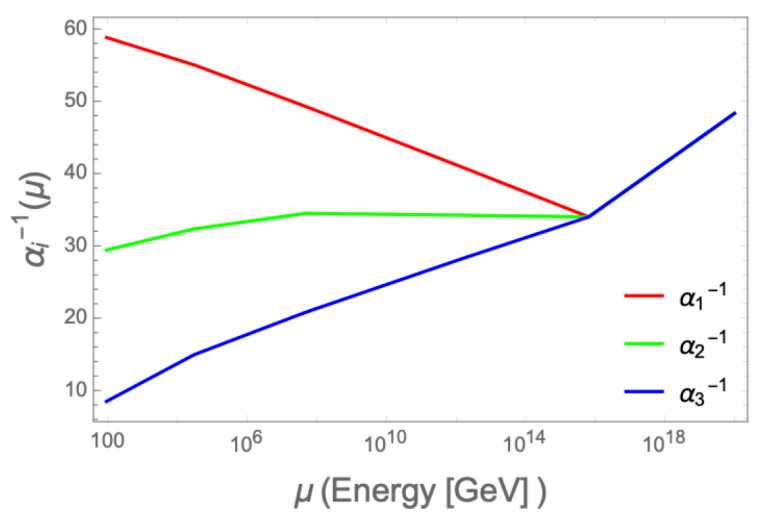}
    \caption{Runnings of the gauge coupling constants, which realize GCU with three light components of {\bf 45} Higgs, are depicted for the mass spectrum (1) in the left panel and (2) in the right panel.}
    \label{fig:GCU2}
\end{figure}

\begin{table}[H]
\caption{Mass spectra of {\bf45} Higgs components which satisfy the GCU condition when the $(\overline{3},3)$, $(8,2)$, $(\overline{3},1)$ components are light.}
\centering
\scalebox{0.75}{
  \begin{tabular}{|l|c|c|c|c|c|c|c|c|c|c|} \hline
  	& $M_{XY}$ & $M_{H_C}$ & $M_{\Sigma}$ & $M(1,2)$ & $M(\overline{3},1)$ & $M(\overline{3},3)$ & $M(3,1)$ & $M(3,2)$ & $M(6,1)$ & $M(8,2)$  \\ \hline
  	$(1)$  & $6.0\times10^{15}$ & $6.0\times10^{15}$ & $6.0\times10^{15}$ & $6.0\times10^{15}$ & $1.0\times10^{13}$ & $5.8\times 10^{7}$ & $6.0\times10^{15}$ & $6.0\times10^{15}$ & $6.0\times10^{15}$ & $4.7\times10^{4}$ \\ \hline
  	$(2)$  & $8.2\times10^{15}$ & $8.2\times10^{15}$ & $8.2\times10^{15}$ & $8.2\times10^{15}$ & $1.0\times10^{12}$ & $5.0\times10^{7}$ & $8.2\times10^{15}$ & $8.2\times10^{15}$ & $8.2\times10^{15}$ & $4.2\times10^{4}$ \\ \hline
  \end{tabular}}
  \label{mass spectra2}
\end{table}
\noindent
In Fig.~\ref{fig:GCU2}, we illustrate the evolution of the gauge couplings for the mass spectra in Table~\ref{mass spectra2}. It is evident from the figures that GCU is successfully realized, thanks to the presence of the three light components.
Later, in Fig.~\ref{region}, we plot the allowed region where GCU is achieved and the proton lifetime bounds are satisfied on the ${\overline M}$-$M(\overline{3},1)$ plane in the scenario involving three light components.

\section{Proton Decay\label{proton decay}}
\begin{figure}[H]
	 \centering
\includegraphics[width=70mm]{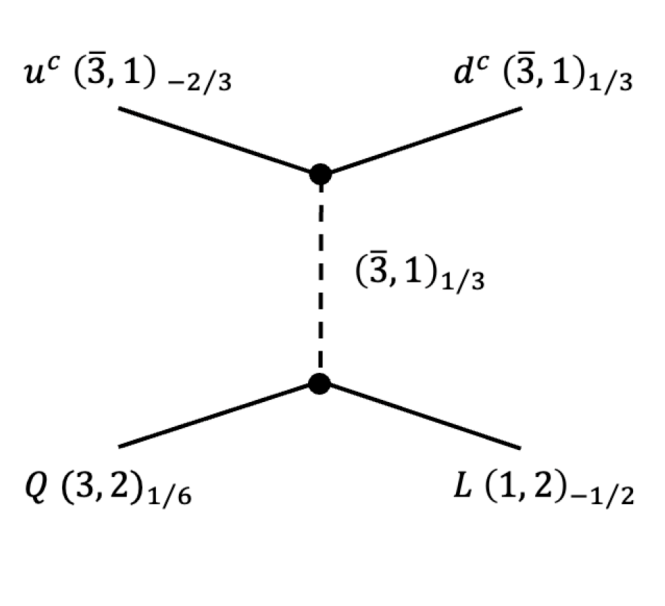} 
\caption{Feynman diagram inducing proton decay via the exchange of $(\overline{3},1)_{1/3}$ in the {\bf 45} Higgs boson.}
	 \label{Proton decay}
\end{figure}

Of the components of the {\bf 45} Higgs field, only $(\overline{3},1)_{1/3}$ contributes to proton decay~\cite{Dorsner:2009cu,Dorsner:2012nq}, as illustrated in Fig.~\ref{Proton decay}.
If the mass of $(\overline{3},1)_{1/3}$ falls below the GUT scale, it may induce rapid proton decay. Thus, it is important to check whether our model leads to excessively rapid proton decay.

In fact, the $(\overline{3},1)_{1/3}$ component of the {\bf 45} Higgs field mixes with that of the {\bf 5} Higgs field.
Here we assume that the mixing terms for the {\bf 45} and {\bf 5} Higgs fields are sufficiently small that this mixing is negligible in the analysis of proton decay.

Now let us estimate the partial widths of proton decay via the exchange of the $(\overline{3},1)_{1/3}$ boson contained in the {\bf 45} Higgs field.
The $(\overline{3},1)_{1/3}$ boson couples to quarks/leptons through the Yukawa coupling of the {\bf 45} Higgs boson $Y_{\overline{45}}$.
Thus we need to estimate $Y_{\overline{45}}$.
To this end, we note that we have assumed that the {\bf 45} Higgs field couples exclusively to the second generation fermions, resulting in the realization of Georgi-Jarlskog-type fermion mass matrices as~\cite{GeorgiJarlskog}
\begin{align}
	M_u=
	\begin{pmatrix}
	0 & D & 0\\
	D & 0 & F\\
	0 & F & E\\
	\end{pmatrix},~~~~
	M_d=
	\begin{pmatrix}
	0 & A & 0\\
	A & C & 0\\
	0 & 0 & B\\
	\end{pmatrix},~~~~
	M_e=
	\begin{pmatrix}
	0 & A & 0\\
	A & -3C & 0\\
	0 & 0 & B\\
	\end{pmatrix},
\end{align}
where we use the same convention as Ref.~\cite{GeorgiJarlskog}.
Additionally, to realize the observed fermion mass hierarchy, these mass matrices should have
the following hierarchical structures:
\begin{align}
	M_u\sim
	\begin{pmatrix}
	0 & \lambda^6 & 0\\
	\lambda^6 & 0 & \lambda^2\\
	0 & \lambda^2 & 1\\
	\end{pmatrix},~~~~
	M_d\sim M_e\sim
	\begin{pmatrix}
	0 & \lambda^3 & 0\\
	\lambda^3 & \lambda^2 & 0\\
	0 & 0 & 1\\
	\end{pmatrix},
\end{align}
where $\lambda\simeq 0.22$ and coefficients of $\mathcal{O}(1)$ are omitted. These mass matrices are diagonalized as 
\begin{align}
	&M_u\sim 
	\begin{pmatrix}
	0 & \lambda^6 & 0\\
	\lambda^6 & 0 & \lambda^2\\
	0 & \lambda^2 & 1\\
	\end{pmatrix}
	\sim
	\begin{pmatrix}
	1 & \lambda^2 & 0\\
	\lambda^2 & 1 & \lambda^2\\
	0 & \lambda^2 & 1\\
	\end{pmatrix}
	\begin{pmatrix}
	\lambda^8 & 0 & 0\\
	0 & \lambda^4 & 0\\
	0 &  0 & 1\\
	\end{pmatrix}
	\begin{pmatrix}
	1 & \lambda^2 & 0\\
	\lambda^2 & 1 & \lambda^2\\
	0 & \lambda^2 & 1\\
	\end{pmatrix},\label{mixingMu}\\
	&M_d\sim M_e
	\sim 
	\begin{pmatrix}
	0 & \lambda^3 & 0\\
	\lambda^3 & \lambda^2 & 0\\
	0 & 0 & 1\\
	\end{pmatrix}
	\sim
	\begin{pmatrix}
	1 & \lambda & 0\\
	\lambda & 1 & 0\\
	0 & 0 & 1\\
	\end{pmatrix}
	\begin{pmatrix}
	\lambda^4 & 0 & 0\\
	0 & \lambda^2 & 0\\
	0 & 0 & 1\\
	\end{pmatrix}
	\begin{pmatrix}
	1 & \lambda & 0\\
	\lambda & 1 & 0\\
	0 & 0 & 1\\
	\end{pmatrix}\label{mixingMdMe}.
\end{align}
Here the diagonalizing matrices on the left and right are identified with the mixing matrices 
 that connect the fermion mass eigenstates and the flavor basis where the {\bf 45} Higgs field couples solely to the second generation fermions. These diagonalizing matrices provide insight into the magnitudes of flavor mixings in interactions involving the $(\overline{3},1)_{1/3}$ boson and quarks/leptons. 
From the diagonalizing matrices obtained in Eqs.~(\ref{mixingMu}),(\ref{mixingMdMe}) and the estimate of $C\sim \lambda^2$,
 the Yukawa coupling matrix $Y_{\overline{45}}$ is estimated as follows:
If the {\bf left side} of $Y_{\overline{45}}$ is in the mass eigenbasis of left-handed and right-handed up-type quarks,
 and the {\bf right side} is in the mass eigenbasis of left-handed charged leptons and right-handed down-type quarks,
 then $Y_{\overline{45}}$ is expressed as
\begin{align}
Y_{\overline{45}}=   
	\frac{v}{v_{45}}
	\begin{pmatrix}
	1 & \lambda^2 & 0\\
	\lambda^2 & 1 & \lambda^2\\
	0 & \lambda^2 & 1\\
	\end{pmatrix}
	\begin{pmatrix}
	0 & 0 & 0\\
	0 & \lambda^2 & 0\\
	0 & 0 & 0\\
	\end{pmatrix}
	\begin{pmatrix}
	1 & \lambda & 0\\
	\lambda & 1 & 0\\
	0 & 0 & 1\\
	\end{pmatrix}=
	\frac{v}{v_{45}}
	\begin{pmatrix}
	\lambda^5 & \lambda^4 & 0\\
	\lambda^3 & \lambda^2 & 0\\
	\lambda^5 & \lambda^4 & 0\\
	\end{pmatrix}.
\end{align}
Here $v=246$~GeV is the VEV of the SM Higgs field, and $v_{45}$ is the portion of $v$ carried by the {\bf 45} Higgs field, namely, the VEV of the {\bf 45} Higgs field is given by $v_{45}$.
If we write the VEV of the {\bf 5} Higgs field as $v_5$, we have $v^2=v_{45}^2+v_5^2$.

If the {\bf left side} of $Y_{\overline{45}}$ is in the mass eigenbasis of left-handed down-type quarks,
 and the {\bf right side} is in the mass eigenbasis of left-handed charged leptons,
 then $Y_{\overline{45}}$ is expressed as
\begin{align}
Y_{\overline{45}}=
	\frac{v}{v_{45}}
	\begin{pmatrix}
	1 & \lambda & 0\\
	\lambda & 1 & 0\\
	0 & 0 & 1\\
	\end{pmatrix}
	\begin{pmatrix}
	0 & 0 & 0\\
	0 & \lambda^2 & 0\\
	0 & 0 & 0\\
	\end{pmatrix}
	\begin{pmatrix}
	1 & \lambda & 0\\
	\lambda & 1 & 0\\
	0 & 0 & 1\\
	\end{pmatrix}=
	\frac{v}{v_{45}}
	\begin{pmatrix}
	\lambda^4 & \lambda^3 & 0\\
	\lambda^3 & \lambda^2 & 0\\
	0 & 0 & 0\\
	\end{pmatrix}.
\end{align}
Utilizing the estimates of $Y_{\overline{45}}$ above, we calculate the partial widths of proton decay via $(\overline{3},1)_{1/3}$ boson exchange. 
We concentrate on six decay modes listed in Table~\ref{proton decay}, since they are relatively severely constrained by experiments.
The partial widths are calculated as~\cite{Nath:2006ut}
\begin{align}
\Gamma(p\to\mu^+K^0)&=\frac{1}{64\pi}\left(1-\frac{m_K^2}{m_p^2}\right)^2\frac{m_p}{f^2} \ (-1-D+F)^2\alpha_H^2A_{RL}^2 \frac{1}{M(\overline{3},1)^4}(\lambda^4\cdot\lambda^4)^2    \left(\frac{v}{v_{45}}\right)^4,
\\
\Gamma(p\to\mu^+\pi^0)&=\frac{1}{64\pi}\left(1-\frac{m_\pi^2}{m_p^2}\right)^2\frac{m_p}{f^2} \ \frac{1}{2}(1+D+F)^2\alpha_H^2A_{RL}^2 \frac{1}{M(\overline{3},1)^4}(\lambda^5\cdot\lambda^4)^2        \left(\frac{v}{v_{45}}\right)^4,
\\
\Gamma(p\to e^+K^0)&=\frac{1}{64\pi}\left(1-\frac{m_K^2}{m_p^2}\right)^2\frac{m_p}{f^2} \ (-1-D+F)^2\alpha_H^2A_{RL}^2 \frac{1}{M(\overline{3},1)^4}(\lambda^4\cdot\lambda^5)^2
\left(\frac{v}{v_{45}}\right)^4,
\\
\Gamma(p\to e^+\pi^0)&=\frac{1}{64\pi}\left(1-\frac{m_\pi^2}{m_p^2}\right)^2\frac{m_p}{f^2} \ \frac{1}{2}(1+D+F)^2\alpha_H^2A_{RL}^2 \frac{1}{M(\overline{3},1)^4}(\lambda^5\cdot\lambda^5)^2
\left(\frac{v}{v_{45}}\right)^4,
\\
\Gamma(p\to \bar{\nu}K^+)&=\frac{1}{64\pi}\left(1-\frac{m_K^2}{m_p^2}\right)^2\frac{m_p}{f^2}\ \alpha_H^2A_{RL}^2 \frac{1}{M(\overline{3},1)^4}
\left\{\lambda^4\cdot\lambda^3 \ \frac{2D}{3} + \lambda^5\cdot\lambda^2\left(1+\frac{D}{3}+F\right)\right\}^2
\nonumber\\
&\ \ \ \ \ \ \ \ \ \ \ \ \ \ \ \ \ \ \ \ \ \ \ \ \ \ \ \ \ \ \ \ \ \ \ \ \ \ \ \ \ \ \ \ \ \ \ \ \ \ \ \ \ \ \ \ \ \ \ \ \ \ \ \ \ \ \ \ \ \ \ \ \times \left(\frac{v}{v_{45}}\right)^4,
\\
\Gamma(p\to \bar{\nu}\pi^+)&=\frac{1}{64\pi}\left(1-\frac{m_\pi^2}{m_p^2}\right)^2\frac{m_p}{f^2} \ (1+D+F)^2\alpha_H^2A_{RL}^2 \frac{1}{M(\overline{3},1)^4}(\lambda^5\cdot\lambda^3)^2\left(\frac{v}{v_{45}}\right)^4,
\end{align}
 where $m_p,m_K,m_\pi$ respectively denote the masses of proton, kaon, pion, $f$ denotes the pion decay constant, $D,F$ are parameters of the baryon chiral Lagrangian, $\alpha_H$ is the hadronic form factor, $A_{RL}$ accounts for renormalization group evolutions of the dimension-six operators,
 and $M(\overline{3},1)$ denotes the mass of the $(\overline{3},1)_{1/3}$ boson.
For $m_p,m_K,m_\pi$, we use the values in Particle Data Group~\cite{pdg}.  We have $D=0.80$, $F=0.46$, $f=0.093$~GeV.
For $\alpha_H$, we use the result of Ref.~\cite{Aoki:2017puj} as $\alpha_H(2~{\rm GeV})=-0.0144$~GeV$^3$. 
We have evaluated $A_{RL}$ by solving 1-loop RGE for the effective coupling of the proton decay operator and found $A_{RL}=2.6$.

We study the constraints on the mass of the $(\overline{3},1)_{1/3}$ boson from the proton partial lifetime bounds set by Super-Kamiokande~\cite{Super-Kamiokande:2020wjk},\cite{Super-Kamiokande:2022egr},\cite{Super-Kamiokande:2014otb},\cite{Super-Kamiokande:2005lev},\cite{Super-Kamiokande:2013rwg}. 
The current experimental bound on partial lifetime and the lower limit of the mass obtained from it for each mode
 are listed in Table~\ref{proton decay}.
Also, we show how the partial widths depend on the factor of $\lambda$.
In this analysis, we fix $v/v_{45}=\sqrt{2}$.
Note that the top quark Yukawa coupling comes solely from the coupling of the {\bf 5} Higgs field, and so it is enhanced by $v/v_5$.
To avoid the top quark Yukawa coupling becoming non-perturbative through RG running below the GUT scale, we need $v/v_5\lesssim \sqrt{2}$, and hence $v/v_{45}\gtrsim \sqrt{2}$.
Therefore, our choice of $v/v_{45}=\sqrt{2}$ leads to the most mild bound on the mass of the $(\overline{3},1)_{1/3}$ boson.

\begin{table}[H]
  \caption{Lower limits of the mass of the $(\overline{3},1)_{1/3}$ boson obtained from the 90\% CL bounds on proton partial lifetimes.}
  \begin{tabular}{|c|c|c|c|c|c|c|}
\hline
   Decay mode & $p\rightarrow\mu^+K^0$ & $p\rightarrow\mu^+\pi^0$ & $p\rightarrow e^+K^0$ & $p\rightarrow e^+\pi^0$ & $p\rightarrow \bar{\nu}K^+$ & $p\rightarrow \bar{\nu}\pi^+$\\ \hline
   Partial width & $\propto(\lambda^8)^2$ & $\propto(\lambda^9)^2$ & $\propto(\lambda^9)^2$ & $\propto(\lambda^9)^2$ & $\propto(\lambda^7)^2$ & $\propto(\lambda^8)^2$ \\ \hline
	90\% CL bound [years]& $3.6\times10^{33}$ & $1.6\times10^{34}$ & $1.0\times10^{33}$ & $2.4\times10^{34}$ & $5.9\times10^{33}$ & $3.9\times10^{32}$  \\ \hline
	Lower limit [GeV]& $1.1\times10^{13}$ & $9.5\times10^{12}$ & $3.7\times10^{12}$ & $4.9\times10^{12}$ & $3.4\times10^{13}$ & $9.5\times10^{12}$ \\ \hline 
    \end{tabular}
  \label{proton decay}
\end{table}
\noindent
From Table~\ref{proton decay}, we find that the $p\rightarrow \bar{\nu}K^+$ mode gives the most stringent bound on the mass of the $(\overline{3},1)_{1/3}$ boson,
 which is $M(\overline{3},1)>3.4\times10^{13}$~GeV.

\begin{figure}[H]
    \centering
     \includegraphics[width=100mm]{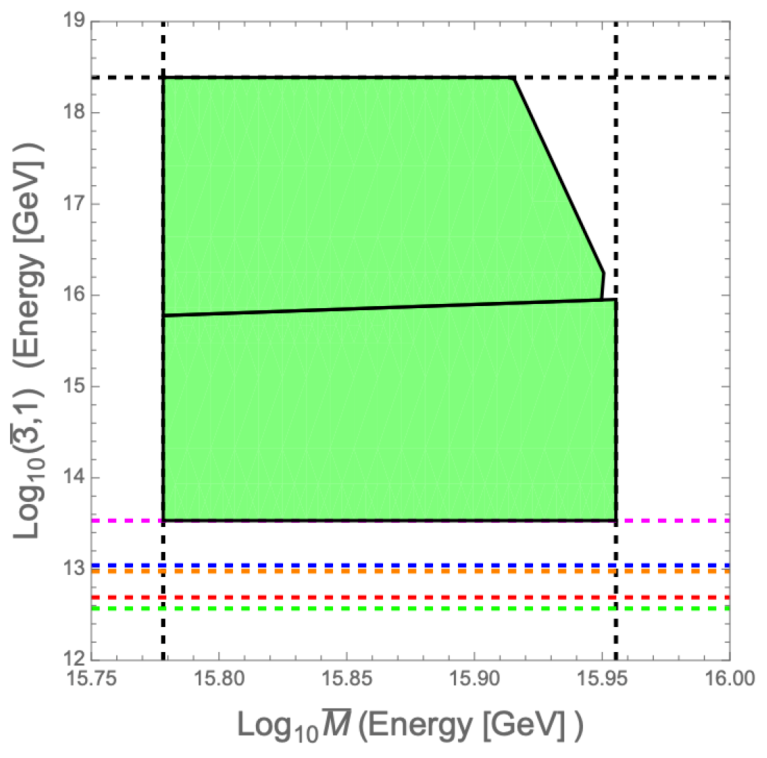}
    \caption{Allowed region on the ${\overline M}$-$M(\overline{3},1)$ plane where GCU is realized and the proton lifetime bounds are satisfied,
    in the scenario where three components of the {\bf 45} Higgs boson are light. The colored region is allowed.
The dashed lines correspond to the bounds on $M(\overline{3},1)$ obtained from various proton decay channels in Table~\ref{proton decay},
  of which the $p\rightarrow \bar{\nu}K^+$ mode gives the most stringent bound of $M(\overline{3},1)>3.4\times 10^{13}$~GeV.
The left side of the colored region is excluded by the constraint from proton decay via the $XY$ gauge boson exchange, while the right side is excluded by the fact that $M(8,2)$ is smaller than the current bound set by the LHC.
}
\label{region}
\end{figure}

In Fig.~\ref{region}, we plot the allowed region where GCU is achieved and the proton lifetime bounds are satisfied, in the scenario with three light components, on the ${\overline M}$-$M(\overline{3},1)$ plane.
The colored region is allowed.
The dashed lines correspond to the bounds on $M(\overline{3},1)$ obtained from various proton decay channels in Table~\ref{proton decay}.
The left side of the colored region is excluded by the constraint from proton decay via the $XY$ gauge boson exchange, while the right side is excluded by the fact that $M(8,2)$ is smaller than the current bound set by the LHC.
\footnote{
We exclude the area above the horizontal line of $M(\overline{3},1)=2.4\times 10^{18}$~GeV because the $(\overline{3},1)_{1/3}$ boson mass exceeds the Planck scale.
}

\section{Conclusion\label{Conclusion}}
We have studied GCU and proton decay in a non-SUSY SU(5) GUT with a {\bf 45} representation Higgs field. Assuming that the components of the {\bf 45} Higgs boson exhibit a hierarchical mass spectrum, we have investigated their contributions to the RGEs in pursuit of GCU. We have found that at least two components of the {\bf 45} Higgs field should have hierarchically small masses to meet the GCU conditions.

In the case where two components of the {\bf 45} Higgs boson are lighter than the other GUT particles, we have observed that GCU is achieved only when the $(\overline{3}, 3)$ and $(8, 2)$ components are light, owing to their substantial contributions to the beta functions.
We have showed the solutions to the GCU conditions as a curve on the $M(\overline{3}, 3)-M(8, 2)$, ${\overline M}-M(8, 2)$, ${\overline M}-M(\overline{3}, 3)$ planes.

In the case where three components of the {\bf 45} Higgs boson are lighter, we have found that the components $(\overline{3},3)$ and $M(8,2)$ tend to be light, similar to the two-light-components scenario. We have focused on the case where the components $(\overline{3},1)$, $(\overline{3},3)$, $(8,2)$ are light, 
because $(\overline{3},1)$ induces proton decay and so we can constrain the model from the proton lifetime bounds. 
We have estimated proton lifetimes under the assumption that the {\bf 45} Higgs boson couples exclusively to the second generation fermions
 so that the fermion mass matrices show the pattern of the Georgi-Jarlskog model.
We have revealed that $p\rightarrow\mu^+K^0$ is the dominant proton decay channel and that the Super-Kamiokande experiment constrains the mass of the $(\overline{3},1)$ component as $M(\overline{3},1)>3.4 \times 10^{13}$~GeV.
We have identified the viable parameter region on the ${\overline M}$-$M(\overline{3},1)$ plane where GCU is achieved and the proton lifetime bounds are satisfied.

\section*{Acknowledgments}
This work is partially supported by Scientific Grants by the Ministry of Education,
Culture, Sports, Science and Technology of Japan, No. 21H00076 (NH) and No. 19K147101 (TY).
The work of TY is supported by Iwanami-Fujukai Foundation.



\end{document}